\documentclass[twocolumn,showpacs,preprintnumbers,amsmath,amssymb,amsfonts,prb,aps]{revtex4}

 \usepackage[dvips]{graphicx}
 \usepackage{dcolumn}
 \usepackage{bm}

\begin{document}


\title{
Diagrammatic quantum field formalism for localized electrons}

\author{S. A. Bonev$^{1}$ and N. W. Ashcroft$^{2}$}
 \affiliation{$^{1}$Department of Physics, Dalhousie University, Halifax, 
                    Nova Scotia B3H 3J5, Canada\\
              $^{2}$Laboratory of Atomic and Solid State Physics,
              Cornell University, Clark Hall, Ithaca, New York 14853-2501}
\date{\today}

\begin{abstract}
We introduce a diagrammatic quantum field formalism for the evaluation
of normalized expectation values of operators, and suitable for
systems with localized electrons. It is used to develop a convergent
series expansion for the energy in powers of overlap integrals of
single-particle orbitals. This method gives intuitive and practical
rules for writing down the expansion to arbitrary order of overlap,
and can be applied to any spin configuration and to any dimension.
Its applicability for systems with well localized electrons has been
illustrated with examples, including the two-dimensional Wigner
crystal and spin-singlets in the low-density electron gas.
\end{abstract}


\maketitle

\newcommand{\vc}[1]{\ensuremath{\mathbf #1}}
\newcommand{\vcr}{\vc{r}}
\newcommand{\aint}[1]{\ensuremath{\int \!\! d\vc{#1}}}
\newcommand{\rhoso}{\ensuremath{\hat{\rho}^{(1)}}}
\newcommand{\rhopo}{\ensuremath{\hat{\rho}^{(2)}}}
\newcommand{\br}[1]{\ensuremath{\langle #1 \rangle}}
\newcommand{\dg}[1]{\ensuremath{#1^{\dagger}}}
\newcommand{\lvac}{\ensuremath{\langle 0 |}}
\newcommand{\rvac}{\ensuremath{|0\rangle}}
\newcommand{\opabc}{\ensuremath{A,B,C,\ldots}}
\newcommand{\prabc}{\ensuremath{A B C\cdots}}
\newcommand{\prabcyz}{\ensuremath{A B C \cdots Y Z}}
\newcommand{\vexp}[1]{\ensuremath{\lvac #1 \rvac}}
\newcommand{\nrm}[1]{\ensuremath{\br{#1|#1}}}
\setlength{\unitlength}{1cm}
\newcommand{\pnt}{\circle{0.1}}
\newcommand{\drl}[3]{\parbox{#1}
                    {\resizebox{#1}{#2}{\includegraphics{#3}}}}
\newcommand{\dcr}[2]{\put(0,0){\pnt} \put(0,0){\vector(#1,#2){0.5}}}
\newcommand{\dan}[4]{\put(0,0){\pnt} \put(#1,#2){\vector(#3,#4){0.5}}}
\newcommand{\pic}[3]{\begin{picture}(#1,#2) #3 \end{picture}}
\newcommand{\pdu}[3]{\pic{#3}{#3}{}
                     \pic{#1}{#2}{\put(0,0){\pnt}\put(0,0){\vector(-1,1){#3}}
                                  \put(-0.35,0.2){\line(1,0){0.3}}
                                  \put(#3,#3){\vector(-1,-1){#3}}
                                  \put(0.35,0.2){\line(-1,0){0.3}} }}
\newcommand{\pdd}[3]{\pic{#3}{#3}{}
                     \pic{#1}{#2}{\put(0,0){\pnt} \put(0,0){\vector(-1,1){#3}}
                                  \put(#3,#3){\vector(-1,-1){#3}} }}
\newcommand{\pmu}[3]{\pic{#3}{#3}{}
                     \pic{#1}{#2}{\put(0,0){\pnt} \put(0,0){\vector(-1,1){#3}}
                                  \put(-0.35,0.2){\line(1,0){0.3}}
                                  \put(#3,#3){\vector(-1,-1){#3}} }}
\newcommand{\pmd}[3]{\pic{#3}{#3}{}
                     \pic{#1}{#2}{\put(0,0){\pnt} \put(0,0){\vector(-1,1){#3}}
                                  \put(#3,#3){\vector(-1,-1){#3}}
                                  \put(0.35,0.2){\line(-1,0){0.3}} }}

%
\section{Introduction}\label{sec_diagr_intro}
%

In the last two decades considerable effort in the theory of electronic
structure has been focused on the development of methods where
the time for computing ground state properties scales linearly with the 
size of the system, referred to as $O(N)$ methods, 
$N$ being the number of electrons in the system.\cite{goe99} 
A standard approach there is to make use of
localized one-particle electron orbitals, and to circumvent their
orthogonalization though various strategies and approximations in 
the subsequent energy
minimization. Indeed, orthogonalization involves computationally 
intense algorithms; it is particularly impractical for geometrical
optimizations, and it is actually intractable when the $N$-body 
electron wave function is to be
written as a linear combination of Slater determinants made from different
single-particle orbitals (i.e. for a general spin state).

The use of non-orthogonal orbitals, on the other hand, poses
its own difficulties, because the antisymmetrization of the many-body 
wave function in this case introduces terms the magnitude of which
increases as $N$, $N^2$, $N^3$, etc 
(leading to the well known orthogonality catastrophe). 
Thus, expectation values of
operators can diverge in the thermodynamic limit, $N
\rightarrow \infty$. It is often the case, in
particular for an arbitrary spin state, that there is no
transparent or/and systematic way of dealing with such problems. 
As a result, the approximation eventually  used may violate, for example,
even the charge neutrality of the system and thus lead to errors that 
also increase with its size.

In this paper, we develop a diagrammatic formalism to deal with
such problems which can be applied for any spin configuration and
in any dimension. We use it to derive a linked cluster theorem for
the evaluation of expectation values (the energy is discussed in
particular) in terms of a convergent series expansion of overlap integrals 
of single-particle orbitals. The diagrammatic language is introduced
by direct analogy with that of standard field theory. The parallel is
indeed interesting, bearing in mind that the case of strongly
localized electrons considered here is the opposite limit of
spatially uniform systems, the traditional domain of many-body 
perturbation theory. The equivalent of the Feynman propagator will be seen
to be the overlap integral, $S$, the single particle orbitals 
correspond to vertices
in the diagrams, and an $n$-body operator introduces $n$ external
points. All diagrams are then calculated in terms of closed loops
connecting the external points. Despite these similarities in
language, however, the linked cluster expansion and the resulting
diagrammatic rules here are quite different from those in 
standard field theory.

Consider now a neutral system
consisting of $N_e$ electrons and $N_i$ ions (or a uniform, positive adn 
rigid background,
in the case of a jellium model) in a volume $V$. The Hamiltonian of
this system is given in atomic units by
\begin{eqnarray} \nonumber 
\hat{H} & = & \sum_{j=1}^{N_e} \frac{\hat{p}^2_j}{2}+
          \sum_{j=1}^{N_i} \frac{\hat{P}^2_j}{2M_i} \\ \nonumber &&
        {} + \frac{1}{2}\aint{r}\aint{r'} \frac{1}{|\vc{r}-\vc{r'}|}\times
        \left[ \hat{\rho}^{(2)}_e(\vc{r},\vc{r'}) \right. \\ && \left.
        {}+ \hat{\rho}^{(2)}_i(\vc{r},\vc{r'})  -
         2\hat{\rho}^{(1)}_e(\vc{r})\hat{\rho}^{(1)}_i(\vc{r'}) \right],
	\label{eq_ham1}
\end{eqnarray}
where the indices $e$ and $i$ refer to electrons and ions,
respectively, $\hat{p}_j$ is the momentum operator of electron
$j$, $\hat{P}_j$ and $M_j$ are the momentum operator and the mass
of ion $j$, and \rhoso and \rhopo are the one- and
two-particle density operators defined respectively by
\begin{equation}
   \rhoso(\vc{r}) = \sum_{j} \delta(\vc{r}-\vc{r}_j),
\end{equation}
and
\begin{equation}
   \rhopo(\vc{r},\vc{r'}) =  \rhoso(\vc{r}) \rhoso(\vc{r'})
                        -  \delta(\vc{r}-\vc{r'})\rhoso(\vc{r'}).
\end{equation}
A standard approach to solving the eigenvalue problem for this system is,
as a first step, to find the solutions of the electronic problem in the
clamped nuclei approximation, where the ionic momenta are set to zero and
their coordinates frozen. The Hamiltonian for this problem is, 
from (\ref{eq_ham1}),
\begin{eqnarray} \label{eq_diagr_ham2}  \nonumber
\hat{H} & = & \sum_{j=1}^{N} \frac{\hat{p}^2_j}{2}+
          \frac{1}{2}\aint{r}\aint{r'}\frac{1}{|\vc{r}-\vc{r'}|}\\ &&
          \times \left[ \hat{\rho}^{(2)}(\vc{r},\vc{r'})
          - 2\hat{\rho}^{(1)}(\vc{r})\rho_b(\vc{r'}) \right]
           + U_b,
\end{eqnarray}
where $\rho_b(\vc{r})$ is the classical density of the positive
charge (ionic or that of a uniform rigid background), $U_b$ is its
self-energy, and we have simplified the notation by dropping the
subscript $e$ from quantities referring to the electrons.

In what follows we discuss the evaluation of the
ground state properties of a system described by the Hamiltonian\
(\ref{eq_diagr_ham2}), and more specifically, the quantity
\begin{equation}\label{eq_diagr_evar}
   E = \frac{\br{\Psi | \hat{H} | \Psi}}{\br{\Psi | \Psi}},
\end{equation}
with $| \Psi \rangle $ an $N$-electron trial state constructed
from localized single-particle spatial orbitals centered at
positions $\{\vc{R}_i\}$. Though (\ref{eq_ham1}) and (\ref{eq_diagr_ham2})
are formally independent of spin, we will nevertheless also allow for
an arbitrary spin configuration (i.e. correlation and even order)
which can be specified by an appropriate set of spin orbitals.

{\em Localized orbitals} here mean that they diminish rapidly away from the
localization centers $\{\vc{R}_i\}$. The limit where space can be divided
into regions each occupied only by a single one-particle function 
corresponds to the semi-classical limit where the spin configuration becomes 
irrelevant and $|\Psi \rangle $ is a product of single-particle states.
When this is not the case, antisymmetrization of the many-body wave 
function and the resulting exchange effects become an important issue in 
determining the structural phase of the ground state.

Moving away from the semi-classical limit, and when the space orbitals are
not orthogonal, requires a necessity to introduce terms in both the 
numerator and denominator
of (\ref{eq_diagr_evar}) that go as $\sim O(S^{2n}N^n$), where $S$ is
an overlap integral between one-electron wave functions, and $n=0,2,3,\ldots$. 
The resulting series are obviously divergent as $N\rightarrow\infty$ 
irrespective of how small but finite $S$ is. 
Here, we will deal with these problems by viewing the overlap effects 
as a formal ``quantum perturbation,'' which introduces some scattering of the
single-particle amplitudes. The normalization of (\ref{eq_diagr_evar})
is then achieved in a diagrammatic approach without an explicit inversion of
an overlap matrix or a requirement to introduce a cut-off radius for 
the localized functions.  The topology of the
connected diagrams that give a convergent and finite expansion for
the energy (per electron) will be determined by the set $\{\vc{R}_i\}$.

The remainder of the paper is organized as follows:
In Section~\ref{sec_diagr_qfn} we summarize a quantum field theoretical
notation, used previously by van Dijk and Vertogen\cite{dve91} and
later by Moulopoulos and Ashcroft\cite{mas93} for describing Wigner crystals. 
All matrix elements relevant for computing the energy are constructed  from products of 
field operators. 
Their anticommutation relations are then used in  Section~\ref{sec_diagr_deme}
to develop a diagrammatic language for evaluating the matrix elements. 
In Section~\ref{sec_diagr_lkt} we show that the taking of a ratio of matrix
elements leads to a linked cluster expansion. First, an algebraic 
expansion is obtained by generalizing a  mathematical device used by 
Abarenkov \cite{aba93} in the context of a valence-bond method. Next, the
new formalism is used to prove rigorously that the expansion is
convergent and is topologically equivalent to linked clusters
of closed-loop diagrams. A recipe and an example for calculating
the energy are presented in Section~\ref{sec_diagr_encal}. Further
applications and uses of the method are discussed in 
Section~\ref{sec_diagr_disc}.

%
\section{Quantum field theoretical notation}\label{sec_diagr_qfn}
%

In the formalism of second quantization (requiring specification of an
initiating set of single-particle states), the kinetic energy and
the density operators in (\ref{eq_diagr_ham2}) can be written in the
forms (atomic units are used throughout):
\begin{equation}\label{eq_diagr_kinop}
   \hat{T} = \sum_{\vc{k},s} \frac{k^2}{2}\dg{c}_{\vc{k},s}c_{\vc{k},s},
\end{equation}
\begin{equation}
   \hat{\rho}^{(1)}(\vc{r}) = \sum_{s}\dg{\psi}_s(\vc{r})\psi_s(\vc{r}),
\end{equation}
and
\begin{equation} \label{eq_diagr_rho2op}
   \hat{\rho}^{(2)}(\vc{r},\vc{r'}) = \sum_{s,s'}
                                \dg{\psi}_s(\vc{r})\dg{\psi}_{s'}(\vc{r'})
                                \psi_{s'}(\vc{r'})\psi_s(\vc{r}),
\end{equation}
where $\dg{c}_{\vc{k},s}$ and $c_{\vc{k},s}$ are respectively creation and annihilation
operators for an electron in a plane wave state with a
wave vector $\vc{k}$ and spin $s$, and $\dg{\psi}_s(\vc{r})$
and $\psi_s(\vc{r})$ are the usual field operators, i.e.
\begin{equation}
   \psi_s(\vc{r}) = \frac{1}{\sqrt{V}}\sum_{\vc{k}} e^{i\vc{k}\cdot\vc{r}}
                                                     c_{\vc{k},s},
\end{equation}
which create and annihilate a Fermion with spin $s$ at position
$\vc{r}$. A general state of the system assumes the form
\begin{eqnarray} \nonumber
   |\Psi\rangle &=& \sum_{s_1, \ldots, s_N}
                  \aint{r}_1 \cdots d\vc{r}_N \;
                  F(\vc{r}_1,s_1; \ldots; \vc{r}_N,s_N) \\ && \times
                  \dg{\psi}_{s_1}(\vc{r}_1) \cdots
                  \dg{\psi}_{s_N}(\vc{r}_N)\rvac.  \label{eq_diagr_gst}
\end{eqnarray}
Here $\rvac$ denotes the vacuum state, and the antisymmetrization
of the wave function is implicitly built into (\ref{eq_diagr_gst})
through the anticommutation relations of the field operators,
namely
\begin{equation}\label{eq_diagr_comp1}
   \{\psi_s(\vc{r}),\dg{\psi}_{s'}(\vc{r'})\} = 
   \delta_{s,s'} \;\delta(\vc{r}-\vc{r'}),
\end{equation}
and
\begin{equation} 
   \{\psi_s(\vc{r}),\psi_{s'}(\vc{r'})\} = 
   \{\dg{\psi}_s(\vc{r}),\dg{\psi}_{s'}(\vc{r'})\} = 0. \label{eq_diagr_comp2}
\end{equation}
In variational terms the problem is to determine the amplitude function
$F$ which minimizes (\ref{eq_diagr_evar}). Because we want to
construct the wavefunction from $N$ single-particle space
orbitals, the choices for $F$ are linear combinations of products
of single particle functions, each product representing a
particular fixed-spin configuration (the standard Hartree-Fock
approximation). For example, the simplest {\it ansatz},
corresponding to a ferromagnetic (FM) state is
\begin{equation}
   F(\vc{r}_1,s_1; \ldots; \vc{r}_N,s_N) =
   \prod_{i=1}^N f_i(\vc{r}_i-\vc{R}_i),
\end{equation}
where $f_i(\vc{r}_i-\vc{R}_i)$ indicates the (normalized for convenience)
wavefunction of an electron localized at some position $\vc{R}_i$.

Next, following van Dijk and Vertogen \cite{dve91}, we introduce the
operators $\dg{d}_i$ and $d_i$, defined by
\begin{equation} \label{eq_diagr_ddef}
   \dg{d}_i = \aint{r}\; \dg{\psi}_{s_i}(\vc{r})\:f_i(\vc{r}-\vc{R}_i),
\end{equation}
which create and annihilate an electron localized at position
$\vc{R}_i$, with a one-particle function $f_i(\vc{r})$, and with
spin $s_i$. A state corresponding to a particular fixed-spin
configuration, can now be written as
\begin{equation} \label{eq_diagr_fixsp}
   |\Phi\rangle = \left( \prod_{i = 1}^N \;
                    \dg{d}_i \right) \rvac,
\end{equation}
and if we label all such states by, say $p$,
a general state of the system can be written as a linear
combination of terms of the form (\ref{eq_diagr_fixsp}), i.e.
\begin{equation}
   |\Psi\rangle = \sum_p C_p |\Phi_p\rangle.
\end{equation}
For example, a state corresponding to spin-singlet pairs of electrons will
be described by: \cite{mas93}
\begin{equation} \label{eq_spsng}
   |\Psi\rangle = \prod_{i=1}^{N/2} \, \left(
                  \dg{d}_{i,1\uparrow} \dg{d}_{i,2\downarrow} -
                  \dg{d}_{i,1\downarrow} \dg{d}_{i,2\uparrow}
                  \right) \rvac .
\end{equation}
Here the up and down arrows explicitly indicate the spin to be
associated with the given operator, and it is clear that the
electrons do not have definite spins but are nevertheless grouped in pairs
where the two electrons of each pair always have antiparallel
spins.

From (\ref{eq_diagr_comp1}),  (\ref{eq_diagr_comp2}),
and (\ref{eq_diagr_ddef}), it is straightforward to derive the
following anticommutation relations for the newly defined creation
and annihilation operators, namely:
\begin{equation} \label{eq_diagr_comd1}
   \{ d_i,\dg{d}_j \}  =  \delta_{s_i,s_j} S(i j),
\end{equation}
and
\begin{equation}  \label{eq_diagr_comd2}
   \{ d_i,d_j \}  =  \{ \dg{d}_i,\dg{d}_j \} = 0
\end{equation}
where $S(ij)$, a key quantity in what follows, is
\begin{equation} \label{eq_diagr_sij}
   S(ij) = \aint{r} \; f_i^*(\vc{r}-\vc{R}_i) f_j(\vc{r}-\vc{R}_j),
\end{equation}
the {\em overlap integral} of two single-particle wavefunctions
centered at $\vc{R}_i$ and $\vc{R}_j$. In addition,
\begin{equation} \label{eq_diagr_compd1}
   \{ \psi_s(\vc{r}),\dg{d}_i \}  =  \delta_{s,s_i} f_i(\vc{r}-\vc{R}_i),
\end{equation}
and
\begin{equation} \label{eq_diagr_compd2}
   \{ \psi_s(\vc{r}),d_i \}  =  \{ \dg{\psi}_s(\vc{r}),\dg{d}_i \} = 0.
\end{equation}
Further, if $f_i(\vc{k})$ is the Fourier transform of $f_i(\vc{r})$, then
for a system of dimensionality $D$,
\begin{equation} \label{eq_diagr_comcd1}
   \{ c_{\vc{k},s},\dg{d}_i \}  =  \frac{(2\pi)^{D/2}}{\sqrt{V}}
   e^{-i\vc{k}\cdot\vc{R}_i}\; f_i(\vc{k})\,\delta_{s,s_i},
\end{equation}
and
\begin{equation} \label{eq_diagr_comcd2}
   \{ c_{\vc{k},s},d_i \}  =  \{ \dg{c}_{\vc{k},s},\dg{d}_i \} = 0.
\end{equation}

%
\section{Diagrammatic evaluation of matrix elements}\label{sec_diagr_deme}
%

Within the formalism of the previous section,
all matrix elements of interest for the computation of the
the energy (\ref{eq_diagr_evar}) assume the general form
\begin{equation} \label{eq_diagr_matel}
   \lvac \prabcyz \rvac,
\end{equation}
where $\opabc,$ etc. are creation and
annihilation operators whose anticommutation relations in terms of 
localized single-particle functions have just been established. We now proceed
to interpret these quantities as a sum of closed loop diagrams in a
language very similar to that of standard field-theoretical and 
many-body methods.\cite{psc95,agd75}

We start by selecting an arbitrary labeling order of all distinct operators
of interest; this can be done without loss of generality. 
Distinctions will be based on the label $i$ for the $d_i$, 
$\vc{r}$ for the $\psi_s(\vc{r})$, and $\vc{k}$ for the $c_{\vc{k},s}$
operators; the spin label will be irrelevant. 
Next, we define a $T$-product of operators, $T(\prabc)$,
which is a product of the operators $\opabc$, but written in such an
order that all annihilation operators are on the left and in descending
order of their labels, all creation operators are on the right of the
annihilation operators and in ascending order of their labels, and
the product is multiplied by $(-1)^P$, where $P$ is the number of
permutations needed to obtain the $T$ product from $\prabc$. For example,
\begin{equation}
   T(d_2 d_3 \dg{d}_1 \dg{d}_4 d_1) =
   (-1)^3 d_3 d_2 d_1 \dg{d}_1 \dg{d}_4.
\end{equation}

Next, we define an $N$-product (normal product) of operators, $N(\prabc)$,
which is a product of the operators $\opabc$, where all creation
operators are on the left of all annihilation operators, and the
product is multiplied by $(-1)^P$, with $P$ being the number of
permutations needed to obtain the $N$-ordering from $\prabc$. For
example,
\begin{equation}
   N(d_1d_2\dg{d}_3) = (-1)^2 \dg{d}_3 d_1 d_2 =
                       (-1)^3 \dg{d}_3 d_2 d_1.
\end{equation}
We can now define a pairing, or a contraction of two operators as
\begin{eqnarray} 
   A^c B^c & = & T(A B) - N(A B) \nonumber \\ \label{eq_diagr_contr}
           & = & \left\{ \begin{array}{r@{,\quad}l}
                               \{A,B\} & \textrm{if $AB$ is $T$-ordered}\\
                               -\{A,B\}& \textrm{if $AB$ is not $T$-ordered},
                                        \end{array} \right.
\end{eqnarray}
and then we have the equivalent of Wick's theorem for our problem.
This states that a  $T$-product
can be expressed as a sum of all possible $N$-products with all possible
contractions, i.e.,
\begin{eqnarray} \label{eq_diagr_wick}
   T(\prabcyz) & = & N(\prabcyz) + N(A^c B^c C \cdots Y Z) \nonumber \\
                 & &{} + N(A^c B C^c \cdots Y Z) + \cdots \nonumber  \\
                 & &{} + N(A^a B^c C^a \cdots Y^b Z^c). \label{eq_diagr_wick}
\end{eqnarray}
The validity of the above relation can be verified by inspection,
but it is also not difficult to prove by induction.

Next, taking the vacuum expectation values of (\ref{eq_diagr_contr}) and
(\ref{eq_diagr_wick}), and using the fact that by the definition of an
$N$-product its vacuum average is zero when the product contains 
any uncontracted operators, we have
\begin{equation}
   A^c B^c = \vexp{T(A B)},
\end{equation}
and
\begin{eqnarray} \label{eq_diagr_tprvac}
  \lefteqn{\!\!\!\vexp{T(A B C D \cdots Y Z)} = } \nonumber \\ & & \!\!\!
       \vexp{T(A B)}\vexp{T(C D)}\cdots \vexp{T(Y Z)} \pm \nonumber \\ 
    & & \!\!\!\vexp{T(A C)}\vexp{T(B D)}\cdots \vexp{T(Y Z)} \pm \cdots
\end{eqnarray}
where the $\pm$ signs correspond to the parity of the permutation of the
operators $A B C \cdots X Y Z$. As a consequence, any matrix element 
of the form (\ref{eq_diagr_matel}) is evaluated in complete analogy with
correlation functions in field theory.

Accordingly, we now develop a diagrammatic description for such
matrix elements. The operators we are dealing with have three
attributes: a label associated with the localization center of a
one-particle function (for the $d_i$ operators these functions are
the $f_i(\vc{r})$'s, for the the $\psi_s(\vc{r})$ operators the
$\delta(\vc{r})$'s, and for the $c_{s,\vc{k}}$ operators, the
$f_i(\vc{k})$'s.); a spin orientation; and every operator is either of a 
creation or annihilation character.
So, we will draw points to represent the set of labels
of the operators (these points can obviously be be arranged to reflect the
actual topology of the set $\{\vc{R}_i\}$), and arrows pointing away from or 
towards them for creation or annihilation operators, respectively. 
In addition, we will indicate the spin with a bar across the arrows 
for spin up operators, resulting in what we will refer to as plus 
and minus arrows. For example,
\begin{eqnarray}
   \dg{d}_{i,\uparrow} \;\;\; = \;\;\;
   \pic{1}{1}{\dcr{0}{1}\put(0.2,-0.2){$i$}\put(-0.15,0.2){\line(1,0){0.3}}}
   & \qquad &
   d_{i,\uparrow} \;\;\; = \;\;\;
   \pic{1}{1}{\dan{0}{0.5}{0}{-1}\put(0.2,-0.2){$i$}
                                          \put(-0.15,0.25){\line(1,0){0.3}}} 
   \\
   \dg{d}_{i,\downarrow} \;\;\; = \;\;\;
   \pic{1}{1}{\dcr{0}{1}\put(0.2,-0.2){$i$}}
   & \qquad &
   d_{i,\downarrow} \;\;\; = \;\;\;
   \pic{1}{1}{\dan{0}{0.5}{0}{-1}\put(0.2,-0.2){$i$}}
\end{eqnarray}
and similarly for the $\psi_s(\vc{r})$ and $c_{s,\vc{k}}$ operators.
To extend the analogy within the language of field theory even further,
we will call the points associated with the $d_i$ operators {\em vertices},
and those associated with the  $\psi_s(\vc{r})$ and $c_{s,\vc{k}}$ operators 
{\em external points}; the reason for this choice will become clear later.

In this construction, a pairing of two operators is represented
by a line connecting the points associated with them, and having a
direction determined by their ordering. When the operators are
$T$-ordered, the lines will follow the arrows of the points they
connect. Also, because the commutation relations of opposite
spin operators are zero, only lines connecting either plus or minus arrows
need be considered. If it is not possible to connect all points in this 
fashion, the corresponding matrix element is zero. This means that if
all operators are present in the product as creation-annihilation
pairs (each point has two arrows, one pointing at it and one away from it),
the resulting non-zero diagrams are a collection of closed loops only.

It is now easy to see that after a couple of permutations the
expectation values of the kinetic energy and density operators
(\ref{eq_diagr_kinop})-(\ref{eq_diagr_rho2op}) can be brought to the form 
(\ref{eq_diagr_tprvac}), where all operators are
present in creation-annihilation pairs but all terms involving pairings 
between the $\psi_s(\vc{r})$ and the $c_{s,\vc{k}}$ operators have canceled 
out. Therefore, all relevant matrix elements can indeed be evaluated as 
the sum of all possible closed-loop diagrams that can be constructed
by connecting {\em all} vertices and external points according to
the rules described above. The value of each diagram is then a product of 
the values of each line connecting two points, and the value of each such 
line is the anticommutation relation of the operators represented by the 
points. In addition, a sign must be associated with each diagram, which is
given by
\begin{equation} \label{eq_diagr_sign1}
         (-1)^{n_e}\prod_l (-1)^{n_l-1}
\end{equation}
where $n_e$ is the number of external points, the product is over all 
(closed) loops in the diagram and $n_l$ is the
number of lines (or points) in each loop; the one-point loop diagrams
obviously have no influence on the sign. The formal proof of
({\ref{eq_diagr_sign1}) is straightforward (e.g. by induction). 

By way of example, and to illustrate the rules derived so far, 
we show the diagrammatic
expansion of the one-particle density, $\br{\Psi|\rhoso(\vcr)|\Psi}$, 
where $|\Psi\rangle$ is an $N$-electron ferromagnetic state; thus,
\begin{eqnarray} \nonumber
   \lefteqn{\br{\Psi|\rhoso(\vc{r})|\Psi} =  \delta(\vcr) - 
                 \br{\Psi|\psi(\vcr)\dg{\psi}(\vcr)|\Psi} } \hspace{1cm} 
                 \\[1.2ex] \nonumber
          &=& \;\drl{0.4cm}{1.2cm}{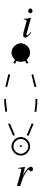}\; - \;
                 \drl{1.2cm}{1.2cm}{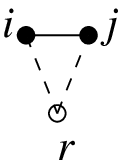}\; -
                \;\drl{0.4cm}{1.2cm}{fig201.eps}\drl{0.4cm}{1.2cm}{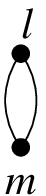}
                  \\ \nonumber
          && {}+\;\;\drl{1.2cm}{1.8cm}{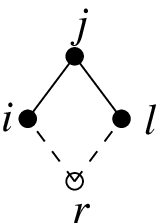}\;\; +
                \;\drl{1.2cm}{1.2cm}{fig204.eps}\drl{0.4cm}{1.2cm}{fig202.eps}
              \;\;-\;\;\drl{1.4cm}{1.8cm}{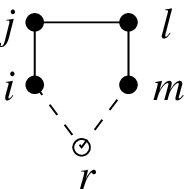} 
                  \\
          &&{}+\;\;\drl{0.4cm}{1.2cm}{fig201.eps}\drl{0.4cm}{1.2cm}{fig202.eps}
                  \drl{0.4cm}{1.3cm}{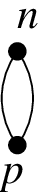}\;\;+- \;\cdots
        \label{eq_merho1}
\end{eqnarray}
Here the filled points indicate summation over all vertices, and we
have omitted one-point loops, which are equal to unity. In
the semi-classical limit, or if the single-particle functions are
orthogonal, only the first diagram above remains. The overlap-order
of a diagram increases, and their values diminish exponentially with
increasing size of the loops. 
The presence of disconnected loops is generally what causes such quantities
to diverge in the thermodynamic limit. As with other many-body methods,
this problem is removed by the normalization of the expectation values, 
which leads to the equivalent of a linked cluster expansion.

%
\section{Construction of a linked cluster expansion} \label{sec_diagr_lkt}
%

Let $|N\rangle$ be a product (or a linear combination of products) of
$N$ creation field operators acting on the vacuum state. As we showed above,
$\br{N | N}$ can be thought of as a sum of all possible closed-loop diagrams 
connecting some representative $N$ points. Then, we can write:
\begin{equation}
   \br{N | N} = \sum\limits_{n_1\cdots n_N} C(n_1,\ldots,n_N),
\end{equation}
where $C(n_1,\ldots,n_N)$ is the class of all diagrams containing
{\em exactly} $n_1$ 1-point loops, $n_2$ 2-point loops, and so on.

Next, let us define a generating function,
\begin{equation}
   Q_N(t) = \sum\limits_{n_1\cdots n_N} C(n_1,\ldots,n_N)\, t^{N-n_1},
\end{equation}
of a standard continuous variable, $t$,
and because  $\sum\limits_{k=1}^N k n_k = N$, it is clear that
\begin{eqnarray}
   Q_N(0) & = &C(N,0,\ldots,0), \\
   Q_N'(0) & = & 0, \\
   Q_N''(0) & = & 2!\,C(N-2,1,0,\ldots,0), \\
   Q_N'''(0) & = & 3!\,C(N-3,0,1,\ldots,0),
\end{eqnarray}
etc., or more generally, for the $m^{\textrm{th}}$ derivative of $Q_N(t)$:
\begin{equation}
   Q_N^{(m)}(0) = m! \sum\limits_{n_2,\ldots,n_N} C(N-m,n_2,\ldots,n_N),
\end{equation}
subject to the constraint $\sum\limits_{k=2}^N\, k n_k = m$.

Furthermore, we can also define a function, associated with $Q_N(t)$, by
\begin{equation}
R_n(t) = \frac{Q_{N+n}(t)}{Q_N(t)}
\end{equation}
and think of the original $N$ points as vertices representing 
one-particle functions, and the additional $n$ points as {\em external}
and representing an $n$-body operator $\hat{O}^{(n)}$. Then, within
this construction,
\begin{eqnarray} \nonumber
   \frac{\br{\Psi|\hat{O}^{(n)}|\Psi}}{\br{\Psi|\Psi}} & = &
                  R_n(1) \\ 
   & & \hspace{-0.5cm} {}- \left\{ \begin{array}{c} 
           \textrm{diagrams with lines}\\
           \textrm{connecting external points}\end{array}\right\}
\end{eqnarray}
and a diagrammatic expansion of the above can be obtained by 
considering the Taylor expansion of $R(t)$ around $t=0$, namely,
\begin{equation} \label{eq_diagr_r1exp}
   R_n(1) = \sum\limits_{m=0}^\infty \sum\limits_{i=0}^m \;
          \frac{1}{i!}Q_N^{(i)}(0) \; \frac{1}{(m-i)!}\,
          \left(\frac{1}{Q_{N+n}(0)}\right)^{(m-i)}.
\end{equation}
For our further discussion it will be convenient to
denote by $V_i$  the value of all diagrams that can be constructed
out of {\em any} $i$ vertices, and that do not contain 1-point loops. 
Similarly, $X_i$ will indicate all such diagrams, but where the $i$ points
{\em may} include external points. Clearly,
\begin{equation}
   V_i = \frac{1}{i!}Q_N^{(i)}(0),
\end{equation}
and
\begin{equation}
   X_i = \frac{1}{i!}Q_{N+n}^{(i)}(0).
\end{equation}

\subsection{Fixed-spin configuration} \label{ssec_fspin}
 
First, we consider the case when $|N\rangle$ is a single product of
$\dg{d}_i$ operators. In such a fixed-spin configuration, $Q_N(0) = 1$, and
$\frac{1}{(m-i)!}\,\left(\frac{1}{Q_{N+n}(0)}\right)^{(m-i)}$ can be 
decomposed simply as
\begin{eqnarray} \nonumber
   \lefteqn{\frac{1}{(m-i)!}\,\left(\frac{1}{Q_{N+n}(0)}\right)^{(m-i)} =  
            W_{m-i} } \hspace{2cm} \\
   & = & 
   \sum\limits_{j_1\cdots j_{m-i}} \; X_{j_1}\cdots X_{j_{m-i}},
    \label{eq_wmmi}
\end{eqnarray}
where $j_1 + \cdots + j_{m-i} = m-i$.
Then, because $R_0(1) = 1$, we have
$\sum\limits_{i=0}^m X_i W_{m-i} = 0$, and it follows that:
\begin{eqnarray} \nonumber \label{eq_diagr_r1fs}
   R_n(1) & = &\sum\limits_{m=0}^\infty \sum\limits_{i=0}^m \;
                                          (V_i - X_i) W_{m-i} \\ \nonumber
        & = & 1 + (V_2 - X_2) + (V_3 - X_3)  \nonumber \\ & & {}
              + \left[ V_4 - X_4 - 6X_2(V_2 - X_2)\right] + \cdots
          \label{eq_diagr_r1fs}
\end{eqnarray}
This expression now represents a {\em convergent} linked cluster
expansion. It is easy to see that the second and third terms above
are simply two-point and three-point loops involving
the external points. The higher order terms are more complicated,
but they are equivalent to {\em chained} loops connected to the external 
points. For example, the fourth term is a sum of all four-point loops
involving external points and a product of two two-point loops, either
chained or not, but connected to two different external points. 
To prove that (\ref{eq_diagr_r1fs}) is indeed convergent and
to illustrate the diagrammatic rules, we present a second construction for 
the linked cluster expansion.

Thus, we will now consider only the case when $n=1$ and
introduce the following notation:
$L_{r i_1\ldots i_m}$ will denote a single loop connecting 
all points labeled $r, i_1,\ldots, i_m$, $|N-\{i_1\ldots i_m\}\rangle$
will be a state obtained by removing operators 
$\dg{d}_{i_1}\ldots \dg{d}_{i_m}$ from $|N\rangle$, $\bar{R}_{i_1\ldots i_m}$
will then be defined by the ratio
\begin{equation} \label{eq_rbardef}
   \bar{R}_{i_1\ldots i_m} = \frac{\nrm{N-\{i_1\ldots i_m\}}}{\nrm{N}},
\end{equation}
and $D_{i_1\ldots i_m}$ will denote the subset of all diagrams from
$\nrm{N}$, which have at least one non-unity loop connected to
any of the points $\{i_1\ldots i_m\}$, divided by $\nrm{N}$. 
Then, $R_1(1)$ can be expanded in the following way:
\begin{equation} \label{eq_r11}
   R_1(1) = L_r 
        + \frac{1}{m!}\sum_{m=1}^N L_{r i_1\ldots i_m}\bar{R}_{i_1\ldots i_m},
\end{equation}
where summation over repeated indexes is implied, and $1/m!$ is to 
take account of repetitions.

To every $\bar{R}_{i_1\ldots i_m}$ term, we now add and subtract 
$D_{i_1\ldots i_m}$ leading to
\begin{eqnarray} \nonumber
   R_1(1) & = & L_r + \frac{1}{m!}\sum_{m=1}^N L_{r i_1\ldots i_m} \\&& {}
                    - \frac{1}{m!}\sum_{m=1}^N L_{r i_1\ldots i_m}
                                                 D_{i_1\ldots i_m}.
            \label{eq_r11it1}
\end{eqnarray}
Subsequently, the terms $D_{i_1\ldots i_m}$ can 
be decomposed into products of loops connected to the points
$\{i_1\ldots i_m\}$ and ratios $\bar{R}_{i_1\ldots i_m\ldots}$.
By repeating this procedure, we are building a {\em chained} structure
of loops connected to the external point, $r$. Every repetition contributes
a minus sign and {\em exactly} one more (surviving) loop to the chain. 
To see the latter, consider a particular element, say
\begin{eqnarray} \nonumber
 \lefteqn{D_{i_1 i_2} =  L_{i_1 i_2}R_{i_1 i_2} 
   + 2\sum_{m=1}^{N-1}L_{i_1 j_1\ldots j_m}R_{i_1 i_2 j_1\ldots j_m}
          } \\ && {} 
   + \sum_{m,n=1}^{N-2}L_{i_1 j_1\ldots j_m}L_{i_2 l_1\ldots l_n}
                       R_{i_1 i_2 j_1\ldots j_m l_1\ldots l_n}.
     \label{eq_di1i2}
\end{eqnarray}
The first two terms above add one, while the third adds 
two loops to the cluster at $i_1$ and $i_2$. However, if 
the recursion procedure is applied to the second term once again, 
it will lead to two sums equal to the third term in (\ref{eq_di1i2}), but
with opposite sign. Thus, each step of the expansion contributes exactly
one loop to the linked cluster that survives subsequent iterations, 
a minus sign to the diagram, and importantly, increases its order by $S^2$.   

The expansion operation, (\ref{eq_r11it1}), can be applied to 
all members of (\ref{eq_r11}) any number of times, $M$, for
any given $N$, until we obtain a sum of all possible loops
involving the external point, $r$, chained to them $0,1,2,3,\ldots$
connected loops (with repetitions) involving the $N$ vertices, and
a remaining leading term of the order 
\begin{equation} \label{eq_rmtor}
   O(S^{2(N+M)})\frac{1}{\nrm{N}} \;\;
   \longrightarrow_{\!\!\!\!\!\!\!\!\!\!\!\!\!\!\!
                    _{N,M\rightarrow\infty}}\;\; 0.
\end{equation}
The construction can be generalized to the case with $n_e$
external points by noting that there will be then simply $n_e$ such linked 
clusters connected to the external points, or alternatively it can be 
seen by considering that $R_n(1) = R^n_1(1)$, e.g.
\begin{equation}
   \frac{\br{N+2|N+2}}{\br{N|N}} = \frac{\br{N+2|N+2}}{\br{N+1|N+1}}
                                   \frac{\br{N+1|N+1}}{\br{N|N}}\, .
\end{equation}

Summarizing to this point, the normalized expectation value of an $n$-body
operator equals the sum of all diagrams where $n=n_e$ external points
are connected by a single loop to linked clusters of loops 
connecting vertices. Loops with lines connecting external points 
directly are not permitted,
while any powers of vertex-only loops are allowed. We already saw that 
an $n_l$-point loop picks up a sign $(-1)^{n_l-1}$. In addition, the 
construction of the linked-cluster expansion shows that the addition
of every new loop alternates the sign, so a diagram with $l$ loops, of which
$l_e$ connect external points, and
$n_e$ external points has to be multiplied also by 
$(-1)^{n_e}(-1)^{l-l_e}$. Altogether the result is that a diagram 
with a total of $N_l$ lines connecting distinct points has a sign
given by
\begin{equation} \label{eq_dsign}
   (-1)^{N_l+n_e-l_e}.
\end{equation}

Continuing with the example of the one-particle density in a single-determinant
many-body state, the normalization of (\ref{eq_merho1}) now gives:
\begin{eqnarray} \nonumber
   \lefteqn{ \rho^{(1)}(\vcr) = 
             \frac{\br{\Psi|\rhoso(\vc{r})|\Psi}}{\nrm{\Psi}}}\hspace{0.9cm}
        \\ \nonumber  
          &=&{} \drl{0.4cm}{1.2cm}{fig201.eps}\; 
                +\; \left\{  \;\drl{1.2cm}{1.2cm}{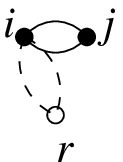}\; -
                \drl{1.2cm}{1.2cm}{fig204.eps}\;       \right\}
        \\ \nonumber
          &&{} \;+\; \left\{ \;\;\drl{1.2cm}{1.8cm}{fig205.eps}\;\;
                 -\;\;\drl{1.2cm}{1.8cm}{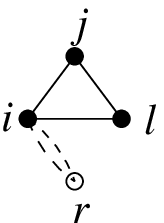}\; \right\}
        \\
          &&{} \;+\; \left\{ \;\;\drl{1.2cm}{1.8cm}{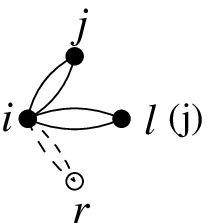}\;\;
                          -\;\;\drl{1.2cm}{1.8cm}{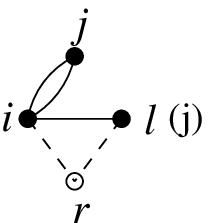}\; \right\}
                + \cdots   \label{eq_rho1nrm}
\end{eqnarray}
Here, as before, filled dots indicate summation over all vertices, 
different labels mark distinct points, and the label $l (j)$ in the
last two diagrams implies that the corresponding point may coincide 
with the point $j$. 
This series is an expansion of the density
\begin{equation} \label{eq_diagr_rhoexp}
   \rho^{(1)}(\vcr) = \rho_0(\vcr) + \rho_1(\vcr) + \rho_2(\vcr) + \cdots,
\end{equation}
where 
\begin{equation}
   \aint{r} \rho_0(\vcr) =  N,
\end{equation}
and for $i\ge 1$
\begin{equation} \label{eq_mpcnd}
   \aint{r} \rho_i(\vcr)  =  0.
\end{equation}
Further, $\rho_0(\vc{r}) = \sum_i |f_i(\vcr-\vc{R}_i)|^2$ is the density 
in the semi-classical limit, or if the one-particle functions
were orthogonal; $\rho_1(\vcr) \equiv 0$; and $\rho_2(\vcr), \rho_3(\vcr),
\rho_4(\vcr),\ldots$ are the terms in curly brackets in (\ref{eq_rho1nrm}),
every one of which represents a different order of overlap.
It is easy to see that they indeed satisfy (\ref{eq_mpcnd}), because
\begin{equation}
    \aint{r}\drl{1.2cm}{1.2cm}{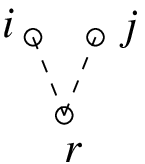}=S(ij)
                    =\;\drl{1.4cm}{0.4cm}{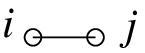}\,.
\end{equation}  
Each of the diagrams in the curly brackets in (\ref{eq_rho1nrm}) represents
a localized effective exchange charge and they can be grouped in pairs forming
electric dipoles. The terms $\rho_i(\vcr)$, for  $i\ge 1$, actually include 
summation over all vertices, and therefore represent higher order multipoles, 
e.g. $\rho_2(\vcr)$ is a quadrupole.

If we now return to the expansion in (\ref{eq_diagr_r1exp}) and compare
it with (\ref{eq_rho1nrm}), we see that $\rho_0(\vcr)$ is given by
$X_2-V_2$. However, the term $X_3-V_3$ gives only the single 3-point loop 
(with a minus sign) in $\rho_2(\vcr)$, so if (\ref{eq_diagr_r1exp}) is 
truncated at this point, charge neutrality in the system will be violated.
The required neutralizing part in $\rho_2(\vcr)$ comes from the next term 
in (\ref{eq_diagr_r1exp}), which contains {\em products} of 2-point loops.
With the diagrammatic formalism, on the other hand, it is intuitively
straightforward to maintain charge neutrality by grouping all diagrams
involving a given set of vertices.

As a second example, the diagrammatic expansion of the two-particle density,
$\rho^{(2)}(\vcr,\vc{r}')$,  consists of the product 
$\rho^{(1)}(\vcr)\rho^{(1)}(\vc{r}')$, which can be obtained from 
(\ref{eq_rho1nrm}), and supplemented by another part with diagrams 
where a single 
loop is associated with both external points, $\vcr$ and ${\vcr}'$. 
Both parts contain overlap-dependent diagrams giving rise to
exchange-correlation effects. Those coming from 
$\rho^{(1)}(\vcr)\rho^{(1)}(\vc{r}')$ arise solely from the non-orthogonality
of the one-particle functions; they are sometimes called {\em indirect
exchange} terms and are usually responsible for the molecular bonding 
(not in a ferromagnetic, but spin-paired state, of course).  
The diagrams where both external points are linked with
a single loop give rise to the so called {\em direct exchange}, and some of them,
for example
\begin{equation} \label{eq_direx}
    (-) \;\;\; \drl{1.3cm}{1.8cm}{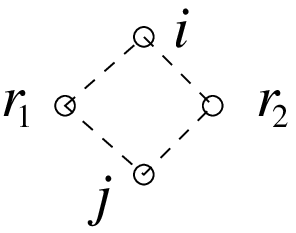}\, ,
\end{equation}
survive even if the one-particle functions are orthogonal. 
Notice that direct exchange comes from parallel spin correlations, and 
indeed, we cannot form a loop such as (\ref{eq_direx}) 
(even with more vertices) so
long as any two electrons in it are in an antiferromagnetic arrangement.
This is not the case for the indirect exchange, where $\vcr$ and 
${\vcr}'$ are in separate loops. The qualitative differences between the 
direct and indirect exchange can also be seen from the fact that
same-order overlap diagrams representing the two terms have opposite
sign (see (\ref{eq_dsign})).  For instance, compare (\ref{eq_direx}) with
\begin{equation}
    (+) \;\;\; \drl{1.3cm}{1.8cm}{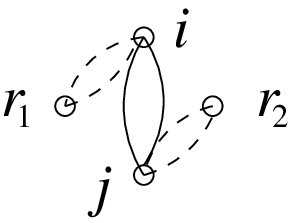}
\end{equation}
Thus, the diagrammatic language accurately captures the
well known fact that the ground state electronic structure is 
often determined by the competition of the two types of exchange.

\subsection{General-spin configuration} \label{ssec_gspin}

Dealing with a general-spin configuration means confronting the fact that 
$|N\rangle$ must be a linear combination of state vectors, each one written 
as a product of $N$ operators,
\begin{equation} \label{eq_genn}
   |N\rangle = \sum_{p=1}^M b_p |N\rangle_p,
\end{equation}
here the $b_p$'s being arbitrary constants. Therefore, we have to consider
$M^2$ different configurations resulting from $\nrm{N}$,
each of them with $N$ points but with
different sets of arrows, representing creation and annihilation operators
and their spins. The difficulties that now arise are related 
first, the fact that a given diagram can be present in more than one 
configuration, and
second, to the consideration that not all diagrams can be constructed 
in all configurations.   
For example, the $C(N,0,\ldots,0)$ class diagrams, 
which are simply (and only) one-point loops, exist only in the $_p\nrm{N}_p$
configurations, and there are $M$ of them. Then, 
\begin{equation} \label{eq_qngen}
   Q_N(0) = \sum_p b_p^2,
\end{equation}
rather than unity, so here we are obliged to keep a tally even of 
the one-point loops.  

As a result of all this, formulating the expansion rules by following the
linked-cluster construction outlined between Eqs.~(\ref{eq_rbardef}) 
and (\ref{eq_rmtor}) might appear to become quite cumbersome for a 
general state because of the 
required book-keeping, even though there are no qualitative differences 
with the single-determinant case.  
However, because the ferromagnetic state leads to a complete set of 
diagrams for a given set of points $\{\vc{R}_i\}$, we can reasonably expect
that the linked-cluster expansion for a general state can be obtained 
from that of a ferromagnetic state (e.g. (\ref{eq_rho1nrm})) by multiplying 
every term in it by a coefficient related to the frequency of occurrence 
of its elements over all spin configurations resulting from $\nrm{N}$. 

This conclusion can be verified by examining Eq.~(\ref{eq_diagr_r1fs}). 
It is still valid when $|N\rangle$ is in the general form (\ref{eq_genn}),
however, the $W_{m-i}$'s, which were previously given by (\ref{eq_wmmi}),
now contain $Q_{N+n}(0)$ to the power $m-i+1$ in the denominator, namely:
\begin{equation}
    W_{m-i}   =  \frac{1}{Q_{N+n}(0)}
        \sum\limits_{j_1\cdots j_{m-i}} \frac{X_{j_1}}{Q_{N+n}(0)}
                                  \cdots \frac{X_{j_{m-i}}}{Q_{N+n}(0)},
\end{equation}
where $Q_{N+n}(0)$ is also given by (\ref{eq_qngen}).
The meaning of the $V_i$'s and $X_i$'s also changes; while in 
Section~\ref{ssec_fspin} they were equal to the two-or-more-point loops
that can be constructed out of $i$ points, now we have to sum over
all configurations coming from $\nrm{N}$ where these same loops
can be formed {\em and} where the remaining $N-i$ points form one-point
loops (i.e. they represent a fixed spin state, $\nrm{N-\{i\}}$). 
In practice, the latter condition actually greatly simplifies the 
calculations, as will be demonstrated in an example below. The final 
result therefore is that the expansion (\ref{eq_diagr_r1fs}), and 
consequently the 
diagrammatic rules derived in  Section~\ref{ssec_fspin}, remain the same 
for the general case, but now every loop carries a coefficient, equal to
\begin{equation}
   \frac{\sum\limits_{\{pp'\}}b_p b_{p'}}{\sum\limits_{p=1}^{M} b_p^2}
\end{equation}
where the sum over $\{pp'\}$ is over all configurations $_p\nrm{N}_{p'}$ 
where (1) the given loop can be formed, and (2) all remaining points are of 
definite spins, i.e. either \pdu{0.5}{0.5}{0.4} or \pdd{.5}{.5}{0.4}.
These coefficients can be thought of as weights of the various loops,
and in the case when all the $b_p$'s are equal to unity, they are
simply the fraction of all configurations in which the given loop 
diagram appears. 

Continuing with the example of the one-particle density, the 
generalization of (\ref{eq_rho1nrm}) is now:
\begin{eqnarray} \nonumber 
   \lefteqn{ \rho^{(1)}(\vcr) =  
                {} \drl{0.4cm}{1.2cm}{fig201.eps}\; 
                +\; c_{ij}\left\{  \;\drl{1.2cm}{1.2cm}{fig207.eps}\; -
                \drl{1.2cm}{1.2cm}{fig204.eps}\;       \right\} }
        \\ \nonumber
          &&{} \;+\; c_{ijl}\left\{ \;\;\drl{1.2cm}{1.8cm}{fig205.eps}\;\;
                 -\;\;\drl{1.2cm}{1.8cm}{fig208.eps}\; \right\}
        \\
          &&{} \;+\; c_{ij}c_{il}\left\{ \;\;\drl{1.2cm}{1.8cm}{fig209.eps}\;\;
                          -\;\;\drl{1.2cm}{1.8cm}{fig210.eps}\; \right\}
                + \cdots   \label{eq_rho1gen} 
\end{eqnarray}
Here, the coefficient associated with the two-point loop connecting
$\vcr$ and $i$ is unity for normalization reasons, and the terms  
in brackets must have the same coefficients in order to preserve
the charge neutrality; both of these statements are actually easy to 
verify explicitly. So, to obtain the expansion to the given order
of overlap for a particular spin state, it is only necessary to 
determine the coefficients for two- and three-point vertex loops,
$c_{ij}$ and $c_{ijl}$ respectively. We will now  show how this is 
done with the example of a spin-singlet paired state (\ref{eq_spsng}).

The wave-function in (\ref{eq_spsng}) is a linear combination of
$M=2^{2/N}$ products of field operators with $b_p=\pm 1$, so
$\sum_p b_p^2 = 2^{N/2}$. If we pick a particular spin pair, 
it leads to four types of configurations in $\nrm{\Psi}$:
\begin{itemize}
   \item[(i)]   $\;$ \pdu{0.5}{0.5}{0.4}   \pdd{0.5}{0.5}{0.4}
   \item[(ii)]  $\;$ \pdd{0.5}{0.5}{0.4}   \pdu{0.5}{0.5}{0.4}
   \item[(iii)] $-$\pmu{0.5}{0.5}{0.4} \pmd{0.5}{0.5}{0.4}
   \item[(iv)]  $-$\pmd{0.5}{0.5}{0.4} \pmu{0.5}{0.5}{0.4}
\end{itemize}
To determine $c_{ij}$, we have to count all configurations where
we can form a two-point loop out of $i$ and $j$, and form one-point
loops of the remaining points. Thus, if $i$ and $j$ belong to the same pair,
they must be either in state (iii) or (iv), thus bringing a factor of $-2$.
The remaining $N-2$ points must be either in configuration (i) or (ii),
of which there are $\frac{1}{2}2^{N/2}$, and all of them with positive sign. 
So, in this case, $c_{ij} = -2\frac{1}{2}2^{N/2}/2^{N/2}=-1$. If $i$ and $j$ 
belong to different pairs, both of these points have to be  
either as in (i) or (ii). This is because the remaining points from
each pair must form one-point loops. From the remaining four combinations
only two survive, because $i$ and $j$ must be associated with parallel spins.
So, the two pairs contribute two configurations, the remaining
$N-2$ points, as before, give rise to $\frac{1}{4}2^{N/2}$ possible
diagrams with only one-point loops, and we therefore find 
$c_{ij} = \frac{1}{2}$.  

For $c_{ijl}$, we have to consider 3-point loops; they can connect either
3 points all belonging to different pairs, or 3 points two of which can 
be from the same singlet pair. In the former case, all points must
be in configurations (i) or (ii), as was the case with $c_{ij}$, and then
they all have to be associated with parallel spins. 
There are two such configurations,
and the remaining $N-3$ points give $\frac{1}{8}2^{N/2}$ more, so the result 
is that $c_{ijl}=\frac{1}{4}$. If 2 of the 3 points belong to the same pair,
they must be either in (iii) or (iv). Then, in either case, the remaining
point must be in either (i) or (ii), but not in both.
So, there are two options each carrying a minus sign. The remaining
$N-2$ points give $\frac{1}{4}2^{N/2}$ eligible combinations, and
in this case  $c_{ijl}=-\frac{1}{2}$.

To summarize, we have determined that for the spin-singlet paired 
state (\ref{eq_spsng}), 
\begin{equation}  \label{eq_cij_1}
  c_{ij} = \left\{ \begin{array}{r@{,\quad}l}
                     -1 & \textrm{if $i$ and $j$ are in the same pair}\\[.2cm]
            \frac{1}{2} & \textrm{if $i$ and $j$ are not in the same pair},
                                        \end{array} \right.
\end{equation}
and
\begin{equation}   \label{eq_cijl_1}
  c_{ijl} = \left\{ \begin{array}{r@{,\quad}l}
     -\frac{1}{2}&\textrm{if any 2 of $i$, $j$, $l$ are in the same pair}
            \\[.2cm]
      \frac{1}{4}&\textrm{if $i$, $j$, $l$ are from different pairs}.
                                        \end{array} \right.
\end{equation}

Result (\ref{eq_cij_1}) is in agreement with Ref's~\onlinecite{mas93} and
\onlinecite{aba93},
but here it is obtained in a quite different way; and with
(\ref{eq_cijl_1}) we are going one step further, as we already have
the next term in (\ref{eq_rho1gen}) without further effort. In fact, the 
expansion (\ref{eq_rho1gen}) has seven terms (if we open the brackets),
however, with the diagrammatic language it is easy to see first, that 
only two of their coefficients are unique, and next to
determine them.

%
\section{Energy calculation} \label{sec_diagr_encal}
%

In this section, we demonstrate the use of the diagrammatic technique for
evaluating the energy of a system with localized electrons. 
First, we formulate general rules for such calculations, and then we 
apply them to a practical example.

\subsection{Diagrammatic rules}

To calculate the energy with the help of the diagrammatic language, we 
adhere to the following procedure:

\begin{enumerate}
  \item Specify the localization points, $\{\vc{R}_i\}$, for the 
        single-particle 
        functions, $f_i(\vcr)$, and decide the required order of overlap.
  \item Determine the coefficients associated with the
        spin configuration for all diagrams up to the required order
        of overlap. The order of overlap of a diagram is usually equal to 
        (but may be higher than) the number of interconnected vertices in it.
  \item Form all connected, topologically non-equivalent and non-zero 
        diagrams with one and two external points up to the required order of 
        overlap following the rules described in Section~\ref{sec_diagr_lkt}.
  \item Determine the signs and symmetry factors (multiplicity) of all 
        diagrams.
  \item Group diagrams involving the same vertices; each group 
        represents either a direct or an indirect (with zero net charge) 
        exchange term. 
  \item With each solid line associate an overlap integral:
           \begin{displaymath}
             \quad  \drl{1.4cm}{0.4cm}{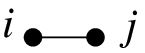} \rightarrow  S(ij) 
                  = \aint{r} f_i(\vcr - \vc{R}_i)f_j(\vcr - \vc{R}_j), 
           \end{displaymath}
        generally assumed to be small.
  \item With each external $k$-point associate a kinetic energy term:
           \begin{displaymath}
               \qquad  \drl{1cm}{1.2cm}{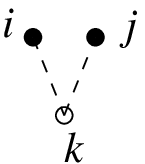} \rightarrow 
                T(ij) = \aint{k} \frac{k^2}{2} f_i(\vc{k})f_j(\vc{k}) \,
                             e^{-i\vc{k}\cdot(\vc{R}_i-\vc{R}_j)}
           \end{displaymath}
  \item With each pair of external points $\vc{r}$ and $\vc{r}'$ associate an 
        interaction energy term:
           \begin{eqnarray*}
             \qquad \lefteqn{\drl{1.5cm}{1.cm}{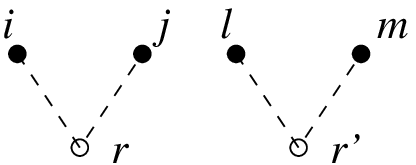} \rightarrow
              U(ij,kl) =} \hspace{1cm}    \\ &&
                          \aint{r}\aint{r'} \frac{1}{|\vcr - \vc{r'}|} 
                          f_i(\vcr)f_j(\vcr)f_l(\vcr')f_m(\vcr'),
           \end{eqnarray*}
        where here $f_i(\vcr)$ stands for $f_i(\vcr-\vc{R}))$, etc.
  \item Sum over all vertex points, $i$,$j$,$l$, and $m$.  
\end{enumerate}

The advantage in following this procedure is that all exchange-correlation
terms originating from the non-orthogonality of the one-particle orbitals
can be easily pre-summed, thus reducing the complexity of the problem to that 
of one with orthogonal orbitals. The computational cost is then limited by 
the efficiency for the evaluation of the Coulomb repulsion integrals 
$U(ij,kl)$. Their computation can be carried out with existing algorithms that 
scale linearly beyond a given $N$, for example, the linear scaling
methods developed by Schwegler and Challacombe \cite{sch99} for computation 
of the $U(ij,kl)$ integrals based on multipole expansions.

\subsection{Example: The two-dimensional Wigner crystal}

As an illustrative example of the application of the procedure described 
above, we consider the case of the ground state of a 2-D Wigner 
crystal\cite{wig34} 
(WC) where the electrons are localized on a hexagonal lattice in the 
presence of a uniform, rigid and neutralizing background. For
$N$ electrons in an area $A$, the background charge density is
$\rho_b=N/A=1/\pi r_s^2$, which also defines the dimensionless 
density parameter $r_s$. Quantum Monte Carlo calculations have  
predicted that the 2-D WC exists for $r_s > 37$.\cite{tce89}
The hexagonal lattice has primitive vectors
\begin{equation}
   \vc{a}_1  =  a(1,0), \qquad \textrm{and} \qquad
   \vc{a}_2  =  \frac{a}{2}(1,\sqrt{3}),
\end{equation}
where $a = \sqrt{\frac{2\pi}{\sqrt{3}}} r_s$ is the lattice parameter, 
and the electrons are localized on lattice sites 
$\vc{R}_i = n_i \vc{a}_1 + m_i \vc{a}_2 $.
With each electron, we associate a normalized Gaussian (trial) 
wavefunction in 2-D with width $\sigma$, i.e.
\begin{equation}
   f(\vcr-\vc{R}_i) = \frac{1}{\sqrt{\pi\sigma^2}} 
                      e^{-(\vcr-\vc{R}_i)^2/2\sigma^2}.
\end{equation}
The choice of Gaussians here is justified because the potential around the
equilibrium positions of the electrons is close to harmonic.\cite{harpt}
Finally, we will restrict our discussion to the antiferromagnetic (AFM)
state - a spin-frustrated structure with alternating lines of up and down 
spins, e.g. an electron localized at  
$\vc{R}_i = n_i \vc{a}_1 + m_i \vc{a}_2 $ will have a positive (negative) 
spin if $m_i$ is even (odd).

With these preliminaries, we can now proceed to calculate the
energy per electron. The overlap integral between
one-particle functions centered at $\vc{R}_i$ and $\vc{R}_j$ is
\begin{equation}
   S(ij) = e^{-R^2_{ij}/4\sigma^2},
\end{equation}
where $\vc{R}_{ij} = \vc{R}_i - \vc{R}_j$. Typical values for $\sigma$ can be
estimated\cite{siges} to be less than $a/4$ and, therefore, the nearest 
neighbor (n.n.) overlap is $S= S(a) \lesssim e^{-1} \approx 0.37$. 
Since $S^4 \approx 0.02$ and $S^5 \approx 0.007$, inclusion of diagrams
up to $O(S^4)$ will guarantee a better than 1\% precision in the calculation of
 the total energy. The next n.n. (n.n.n.) distance in the triangular lattice 
is $\sqrt{3}a$, which means that the n.n.n. overlap integral is 
$S(\sqrt{3}a) = S^3$. Therefore, for the required precision we need to 
consider only n.n. overlaps because the two-vertex diagrams are of order
$S^2(\sqrt{3}a)=S^6$ and the three-vertex are of order 
$S(a)S(\sqrt{3}a)S(a)=S^5$ when they involve n.n.n. overlaps.

The relevant coefficients associated with the spin configuration are
\begin{equation}  \label{eq_cij}
  c_{ij} = \delta_{s_i,s_j},
\end{equation}
and
\begin{equation}   \label{eq_cijl}
  c_{ijl} = \delta_{s_i,s_j}\delta_{s_i,s_l}\delta_{s_j,s_l},
\end{equation}
where $s_i$ indicates the spin of an electron localized at $\vc{R}_i$. 
For diagrams involving only n.n. overlaps, i.e. when 
$R_{ij}=R_{il}=R_{jl}=a$, we have $c_{ijl} = 0$, and we can also set:
\begin{equation}  \label{eq_cij}
  c_{ij} = \left\{ \begin{array}{r@{,\quad}l}
                     1 & \textrm{If $\vc{R}_{ij}=\vc{a}_1$}\\[.2cm]
                     0 & \textrm{if $\vc{R}_{ij}\ne \vc{a}_1$}.
                                        \end{array} \right.
\end{equation}
Thus, the three-vertex $O(S^3)$ diagrams vanish, and we are left with only
two-vertex diagrams of order $S^2$ and $S^4$.

The relevant diagrams with one external point, together with their
signs and multiplicity factors are:
\begin{eqnarray} \label{eq_ex_k0}
O(S^0): & & \;\;\;\drl{0.4cm}{1.2cm}{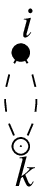} \\ \label{eq_ex_k2}
O(S^2): & & \;\;\drl{1.2cm}{1.2cm}{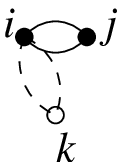}\hspace{0.6cm},\;\;\; 
            -\;\;\drl{1.2cm}{1.2cm}{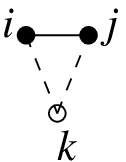} \\   \label{eq_ex_k4}
O(S^4): & & 3\;\drl{1.6cm}{1.2cm}{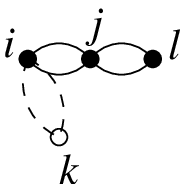}\;,\;\;\; 
           -3\;\drl{1.6cm}{1.2cm}{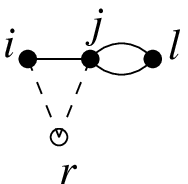}
\end{eqnarray}
Each of the $O(S^2)$ diagrams above has in principle a symmetry factor 
of two -- in the first diagram the external point can be connected to either
$i$ or $j$, and in the second the three-point loop can go either
clockwise or counterclockwise. However, this symmetry factor is taken care of
when performing a sum over $i$ and $j$ and allowing repetition, e.g. 
$\{i,j\}=\{1,2;2,1\}$ (but $i\ne j$). The $O(S^4)$ diagrams have a multiplicity
of three here, because in the triangular lattice there are three 
diagrams of each type, namely, in addition to those shown above, 
when $l=i$, and with a $il$ loop instead of $jl$ 
(for a different spatial or spin configurations we may
have to write these explicitly).

Given (\ref{eq_ex_k0})-(\ref{eq_ex_k4}), we associate the 
following kinetic energy terms:
\begin{eqnarray*}
   \lefteqn{T(ii) + T(ii)S^2(ij)-T(ij)S(ij) +} \hspace{1cm}\\ &&
     3\left[T(ii)S^2(ij)-T(ij)S(ij)\right]S^2(jl),
\end{eqnarray*} 
which with the choice of Gaussian wavefunctions have a simple analytical 
form, namely
\begin{equation}
   T(ij) = S(ij)T(0)\left(1-\frac{R^2_{ij}}{4\sigma^2}\right).
\end{equation}
Here $T(0) = 1/2\sigma^2$ is just the energy of a 2-D harmonic oscillator. 
Then, after summing over $i$, $j$, and $l$, we obtain the kinetic energy 
per electron as:
\begin{equation}
   \frac{T}{N} = T(0)\left[ 1 + 
                            \frac{a^2}{2\sigma^2}(S^2 + 3S^4)\right]\, .
\end{equation}

The relevant diagrams with two external points, representing the 
electron-electron interaction energy are:
\begin{eqnarray}  \label{eq_ex_r0}
O(S^0): &&           \frac{1}{2}\; \drl{0.35cm}{1.1cm}{fig201.eps} \; 
                          \drl{0.35cm}{1.1cm}{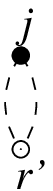} \\[0.4cm]
O(S^2): &&           \drl{.9cm}{1.1cm}{fig207.eps} \label{eq_ex_r2}
                     \drl{0.35cm}{1.1cm}{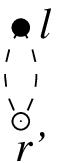} \;,\;\;\;\; 
               - \!  \drl{0.9cm}{1.1cm}{fig204.eps}  \;
                     \drl{0.35cm}{1.1cm}{fig311.eps} \;,\;\;\;\; 
       -\frac{1}{2}\;\drl{1.1cm}{1.2cm}{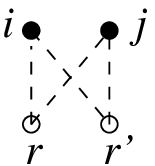} \\[0.4cm]        
O(S^4): &&        3\;\drl{1.2cm}{1.1cm}{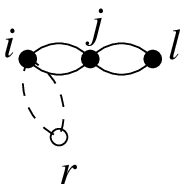} \;  \label{eq_ex_r4}
                     \drl{0.45cm}{1.1cm}{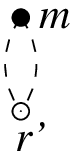} ,\;\;\;
                -3\; \drl{1.2cm}{1.1cm}{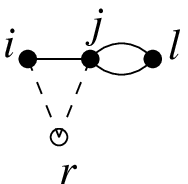} \;
                     \drl{0.45cm}{1.1cm}{fig316.eps}, \\[0.2cm] \nonumber
  &&\hspace{-.0cm} \frac{1}{2}\; \drl{0.9cm}{1.1cm}{fig207.eps} 
                     \drl{0.9cm}{1.1cm}{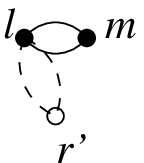}\!\!\!\! ,\;\;\;\;    
       \frac{1}{2}\; \drl{0.9cm}{1.1cm}{fig204.eps}
	             \drl{0.9cm}{1.1cm}{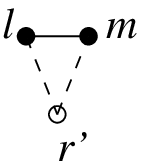}\!\!\!\! ,\;\;\;  
                -\!  \drl{0.9cm}{1.1cm}{fig207.eps}
                     \drl{0.9cm}{1.1cm}{fig314.eps}\!\!\!\!,\\[0.2cm] \nonumber
  &&    -3\frac{1}{2}\;\drl{1.5cm}{1.2cm}{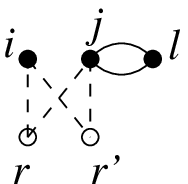} \;
\end{eqnarray}

The factor of 1/2 comes from the symmetry with respect to exchanging $\vcr$ and
$\vc{r'}$, as in Eq. (\ref{eq_diagr_ham2}), and it takes care of overcounting.
 Notice also that the $O(S^0)$ 
term is the Hartree interaction, while the last terms in the  $O(S^2)$ and 
$O(S^4)$ expansions are the direct exchange. The remaining terms  
originate from the product of one-particle density expansions, 
$\rho^{(1)}(\vcr)\rho^{(1)}(\vc{r'})$ and represent multipole interactions. 
With the above 
diagrams we now associate matrix elements $U(ij,lm)$, the general
form of which can be simplified by substituting for the 
$f(\vcr)$ functions and $1/|\vcr - \vcr'|$ their Fourier transforms:
\begin{eqnarray*}
  \lefteqn{ U(ij,lm) = } \hspace{0.3cm} \\
    && \frac{1}{2\pi}\aint{k}_1\!\!\aint{k}_2\!\!\aint{k}_3\!\!\aint{k}_4\,
                  f(\vc{k}_1)f(\vc{k}_2)f(\vc{k}_3)f(\vc{k}_4) \nonumber \\
            && \times \frac{ e^{ i(\vc{k}_1\cdot\vc{R}_i+\vc{k}_2\cdot\vc{R}_j
                                +\vc{k}_3\cdot\vc{R}_m+\vc{k}_4\cdot\vc{R}_m)}}
                           {|\vc{k}_3 + \vc{k}_4|}
                      \delta(\vc{k}_1\!+\!\vc{k}_2\!+\!\vc{k}_3\!+\!\vc{k}_4). 
			   \nonumber
\end{eqnarray*}
Then, changing the integration variables according to: $\vc{k}_1 = -\vc{k}$,
$\vc{k}_2 = \vc{k}-\vc{q}$, $\vc{k}_3 = -\vc{k}'$ and 
$\vc{k}_2 = \vc{k}'+\vc{q}$, and using 
\begin{displaymath}
   \aint{k} f(\vc{k}) f(\vc{k}\pm\vc{q}) e^{-i\vc{k}\cdot\vc{R}_{ij}}
   = S(ij)e^{\sigma^2 q^2 /4} e^{\mp i \vc{q}\cdot\vc{R}_{ij}},
\end{displaymath}
we obtain
\begin{eqnarray}
   \lefteqn{ U(ij,ml) = } \hspace{1cm} \\ &&
                    S(ij)\,S(lm)\;U\left(\frac{\vc{R}_i + \vc{R}_j}{2} - 
                          \frac{\vc{R}_m + \vc{R}_m}{2}\right), \nonumber
\end{eqnarray}
where here
\begin{equation} \label{eq_uij_int}
   U(\vcr) = \sqrt{\frac{2}{\pi\sigma^2}}\int_0^{\pi/2} \!\!d\varphi 
                   e^{-(r^2/2\sigma^2)\cos^2 \varphi}
\end{equation}
is the interaction energy between two Gaussian unit charges with 
centers separated by a distance $r$.

After summing (\ref{eq_ex_r0})-(\ref{eq_ex_r4}) over $i,j,l,$ and $m$, 
the electron-electron interaction energy per electron is then given by:
\begin{eqnarray} \label{eq_vee_os4}
\frac{V_{ee}}{N} & = & \frac{1}{2} \sum_{\vc{R} \ne 0} U(\vc{R})  \\
          &&  + \; (2S^2+10S^4) \sum_{\vc{R} \ne 0}
           \left[ U(\vc{R}) - 
                  U\left(\vc{R}+\frac{\vc{a}_1}{2}\right) \right]\nonumber \\ 
          &&+\;\;\; S^2\left[2\, U\left(\frac{\vc{a}_1}{2}\right)
                         - U(0)\right]\nonumber \\
          &&+\; 4S^4\left[ 10\, U\left(\frac{\vc{a}_1}{2}\right)
                          -4\,U\left(\vc{a}_1\right)- 3\,U(0)\right],\nonumber 
\end{eqnarray}
where terms involving $U(\vc{a}_1)$ and  $U(\vc{a}_1/2)$ have been added and
subtracted in order to complete the second sum above, and 
$U(0)=\sqrt{2\pi}/\sigma$ comes from the direct exchange. 
For the total interaction energy, the 
electron-background and background-background energies have to be added,
which together with the first term in (\ref{eq_vee_os4}) can be 
evaluated by the Ewald lattice summation method. The second sum is 
equivalent to the interaction energy of an ionic lattice with opposite 
charges at $\vc{R}$ and $\vc{R}+\vc{a}_1$ and again is straightforward to 
obtain by the Ewald construction. Finally, the remaining terms require 
only the numerical computation of two integrals such as given by 
(\ref{eq_uij_int}).
  
The solution thus obtained straightforwardly here, up to and including 
fourth order in overlap,
is to be compared with Refs.~\onlinecite{dve91,mas93,aba93}, where similar
problems are discussed but only up to $O(S^2)$. The procedure can easily 
be extended, if needed, to higher orders.

%
\section{Discussion and further examples} \label{sec_diagr_disc}
%

As with other diagrammatic techniques, the benefits here come from a 
translation of an algebraic formalism into a more intuitive diagrammatic 
language. It provides insight helpful for dealing with various spin correlations
and overlap effects of arbitrary order. The diagrammatic rules also
offer guidance for calculating normalized matrix elements in a most efficient 
way for a desired accuracy, while also preserving charge neutrality in the 
system. Violation of charge neutrality as a result of an approximate 
treatment of overlap effects may become a serious issue depending on 
the system size. We will illustrate this problem with an example of a system
of spin singlets, which has been studied previously in the context
of a low-density electron gas,\cite{mas93,aba93,mas92} but also has 
relevance for molecular systems. 

Consider therefore a collection of $n$ spin-singlet pairs of 
electrons ($N=2n$). 
We will assume that the separations between pairs are sufficient so 
that interpair overlaps can be ignored. Without loss of generality we 
will also set all intrapair separations to be the same and equal to $a$ 
(the relevant overlap integral being $S=S(a)$). 
For simplicity, we will examine only the exchange corrections to the 
kinetic energy; these can be easily determined to all orders of 
intrapair overlap. With this construction, there are three types 
of diagrams relevant for the kinetic energy expansion. They are given
by (\ref{eq_ex_k0}) and (\ref{eq_ex_k2}), with the only difference being
that the sign of the $O(S^2)$  diagrams must be changed, since the
corresponding coefficients $c_{ij}$, as given by (\ref{eq_cij}), are
equal to $-1$. To obtain the expansion to all orders of $S$, we have to 
multiply (\ref{eq_ex_k2}) by closed $ij$ loops of all powers resulting in a 
geometric progression. The kinetic energy per electron is thus 
given by:
\begin{eqnarray} \label{eq_ex_tcorr}
   \frac{T}{N} & = & \left[ T(0) + ST(a)\right] (1-S^2+S^4-S^6 + \cdots)
                     \nonumber \\
               & = & \frac{T(0) + ST(a)}{1+S^2} \, .
\end{eqnarray} 

If on the other hand we decide to first truncate the expansions of 
$\br{\Psi|\hat{T}|\Psi}$ and $\br{\Psi|\Psi}$ to a particular order, say 
$S^2$, and then compute $T/N$, the result is:
\begin{equation} \label{eq_ex_tinc}
   \frac{T}{N}  = \frac{T(0) + (n-1)S^2T(0) + ST(a)}{1 + nS^2}\, .
\end{equation}
The difference between (\ref{eq_ex_tcorr}) and (\ref{eq_ex_tinc}) is:
\begin{equation}
  \frac{\delta T}{N} = (n-1)S^4 \frac{T(0)-T(a)/S}{1 + (n+1)S^2 + nS^4}\, ,
\end{equation}
which shows that if $(n-1)S^2 \sim 1$, the error in (\ref{eq_ex_tinc}) 
is comparable to the leading order exchange term. In fact, if 
$n\rightarrow\infty$, (\ref{eq_ex_tinc}) gives $T=NT(0)$, i.e. 
100\% error in the exchange energy. Even if the analysis is carried
for a central pair and only its nearest neighbors, in a typical 
crystalline arrangement $n\sim 10$, so that there is a very stringent
limitation on the allowed overlap, namely $S^2\ll 1/n \sim 0.1$. 

The diagrammatic formalism can also be used to examine the efficiency of 
dealing with the 
non-orthogonality problem by introducing a cut-off radius, $R_c$, 
for the one-particle functions, so that $f(\vcr) = 0$ if $r>R_c$.
If the desired accuracy is second order in overlap, we know that only energy
terms corresponding to two-vertex diagrams such as
\begin{equation} \nonumber
   \drl{0.6cm}{0.3cm}{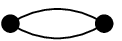}
\end{equation}
need to be considered. However, with a cut-off approach, even if  $R_c$ is 
chosen smaller than the
nearest neighbor distance, one end up calculating (explicitly or implicitly) 
terms of higher
than the required order in overlap, corresponding to diagrams such as:
\begin{equation}  \nonumber
    \drl{1.8cm}{0.3cm}{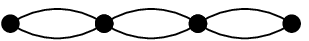}.
\end{equation}
If the profile of the wavefunctions requires larger cut-off radius,
the efficiency would diminish even further as terms corresponding to
3-point, 4-point,etc. loops, would now enter the calculations.

As a final example of application of the diagrammatic formalism, 
we will use it to gain insight into the physics underlying the 
linear scaling density functional theory developed by 
Mauri, Galli and Car,\cite{mgc93,mgg94,gga00} and 
by Ordej\'{o}n {\it et al.}.\cite{odg93}
In this approach, non-orthogonal one particle functions are used and
the inverse of the overlap matrix, $\mathbf{S^{-1}}$, entering the energy 
functional is replaced by a truncated series expansion:
\begin{equation}  \label{eq_sinv}
   \mathbf{S^{-1}} \approx \mathbf{Q} = \sum_{n=0}^{M} (\mathbf{I}-\mathbf{S})^n\, ,
\end{equation}
where $\mathbf{I}$ is the identity matrix, and $\mathbf{S}$ 
has components $S(ij)$. In addition, the following term is added to the energy functional:
\begin{equation} \label{eq_extp}
   \eta \left( N - \int d\vcr \tilde{\rho}(\vcr) \right) \; ,
\end{equation} 
where $\eta$ is a parameter that can be freely chosen, $N$ is the number of electrons 
in the system and $\tilde{\rho}(\vcr)$ is the charge density computed with the truncated
series expansion, Eq. (\ref{eq_sinv}).
This method does not require explicit orthogonalization; a 
minimization of the energy functional of the non-orthogonal Kohn-Sham 
orbitals naturally leads also to orthogonalization. 

Previously, the minimization procedure has been shown to be convergent when the expansion
(\ref{eq_sinv}) is truncated at $M$ odd and $\eta$ is chosen to be positive. 
We can easily see the physical reason for this. In the diagrammatic language, 
an expansion of (\ref{eq_sinv})  to odd $M$
corresponds to considering only  diagrams for the density expansion
where the maximum number of solid lines (representing
overlap integrals) is also odd. The expansion of the density, given by
(\ref{eq_rho1nrm}), shows that truncating the series (\ref{eq_sinv}) in this 
way introduces an error, which is equivalent to decreasing the electron charge density 
and the system becoming not neutral. The extra term added to the energy
functional, (\ref{eq_extp}), then represents the interaction energy 
between a positive external field and net positive charge. 
Thus, reducing this interaction energy to zero, i.e. 
energy minimization, is only achieved when orthogonality is attained.

With this physical picture in mind, it is easy to see that the method should also
work when $M$ is chosen to be odd and the parameter $\eta$ negative. Indeed, in this case the 
error introduced by the truncated expansion (\ref{eq_sinv}) leads to increasing, not decreasing,
electron charge. But with $\eta < 0$, this excess charge now interacts with a negative
field and (\ref{eq_extp}) is again positive definite. Realizing this without the physical
picture in mind is not straightforward because the quantity  
$(\mathbf{Q} - \mathbf{S^{-1}})$, which is negative definite when $M$ is odd 
(see Ref.~\onlinecite{mgc93} for details), 
is not positive definite when $M$ is even. 

We note that for an infinite periodic system with a net charge, 
the long-range Coulombic potential would in principle lead to divergent energy. 
In practice, the divergence can be removed by setting the $\vc{q}=0$ Fourier component of 
the interaction energy to zero -- this is equivalent to adding a uniform potential and 
does not lead to structural changes. The remaining part of the interaction
energy coming from the artificial net charge will be small compared to  (\ref{eq_extp})
if $\eta$ is chosen sufficiently large, 
and will also vanish when orthogonalization is attained.

The above discussion illustrates the utility of the diagrammatic 
formalism to inspect charge neutrality; it is ensured with a proper grouping 
of diagrams, as  shown in Eq.~(\ref{eq_rho1nrm}). 

%
\section{Conclusion} \label{sec_diagr_concl}
%

We have introduced a diagrammatic formalism for the calculation of
normalized expectation values in terms of convergent series 
expansions in powers of one-particle overlap integrals. 
 It can be applied to any order of overlap and for any
spin configuration.
The formalism has been introduced by analogy with conventional
field theoretical methods; however it is applicable for systems with
well localized electrons. As a particular example, we have demonstrated
energy calculations up to fourth order in overlap at the level of 
unrestricted Hartree-Fock and the valence-bond methods. The formalism
presented here can give useful physical insight for the validity of other 
approaches, and potentially be used improve their efficiency.

A possible extension of the formalism can include an analogy of skeletal
diagrams and Dyson-like equations. This would be particularly useful in 
cases where there is a significant overlap among groups of electrons. 
In such cases, selected diagrams, accounting for the overlap among these
electrons, could be summed to an infinite order. This possibility is 
demonstrated with the example from the preceding section, 
Eq.(\ref{eq_ex_tcorr}).

The formalism can also be readily applied for localized bosons. The only 
difference with the fermionic case is in the sign of the diagrams as expected. 
For bosons,
all loops carry a positive sign as a result of the commutation relations, 
however, in the construction of the linked-cluster expansion each chained 
loop still brings a negative sign. Therefore, in this case the sign of a 
diagram is given by $(-1)^{l-l_e}$ instead of (\ref{eq_dsign}), 
where $l-l_e$ is the number of 
closed loops not connected to external points.

%
\section{Acknowledgments} \label{sec_diagr_concl}
%
This work was supported by the National Science Foundation under 
Grant Nos. DMR-9988576, DMR-0302347, and DMR-0601461. 
S.A.B. acknowledges support from the National Sciences and Engineering Research 
Council of Canada.

%


\begin{references}
\bibitem{harpt}  Gauss's theorem is not valid in 2-D, so it is not 
                 immediately apparent that the potential around the 
                 equilibrium positions is close to harmonic. But it can be 
                 seen if one considers a disk of positive
                 uniform charge with radius $r_s$ around each electron. The
                 electrostatic potential of such a disk at $r<r_s$ is 
                 $(4/\pi) r_s E(r/r_s)$, where $E$ is the complete elliptic 
                 integral of the second kind. The potential at $r>r_s$ is
                 $4r_s\left[(r/r_s) E(r/r_s) - (r/r_s - r_s/r)K(r_s/r)\right]$,
                 where $K$ is the complete elliptic integral of the first kind.
                 If the elliptic integrals are expanded in power series, it 
                 can be established that close to the equilibrium position of
                 a given electron, the potential due to the remaining 
                 electrons and all disks is approximately harmonic.
\bibitem{siges}  If the electrons are treated classically in the harmonic 
                 approximation with the potential obtained as outlined 
                 in [\onlinecite{harpt}], 
                 one obtains that $\sigma/r_s \approx r_s^{-1/4}$.
\end{references}
\end{document}